\documentclass[review]{elsarticle}

\usepackage{lineno,hyperref}
\modulolinenumbers[5]

\usepackage{graphicx}
\usepackage{bm}
\usepackage{verbatim}
\usepackage{amsmath}
\usepackage{amssymb}
\hypersetup{colorlinks=false}
\DeclareMathOperator{\tr}{Tr}

\journal{Physics Letters A}

\begin{document}

\begin{frontmatter}

\title{Mean-field solution of the Blume-Capel model under a random crystal field}


\author{P. V. Santos\corref{mycorrespondingauthor}}
\cortext[mycorrespondingauthor]{Corresponding author}
\ead{priscilavs65@yahoo.com.br}

\author{F. A. da Costa}
\ead{fcosta@dfte.ufrn.br}

\author{J. M. de Ara\'{u}jo }
\ead{joaomedeiros@dfte.ufrn.br}

\address{Departamento de F\'{i}sica Te\'{o}rica e Experimental \\ Universidade Federal do Rio Grande do Norte \\ Natal-RN, 59078-900, Brazil}

\begin{abstract}
In this work we investigate the Blume-Capel model with infinite-range ferromagnetic interactions and under the influence of a quenched disorder - a random crystal field.  For a suitable choice of the random crystal field the model displays a wealth of multicritical behavior, continuous and first-order transition lines, as well as re-entrant behavior. The resulting phase diagrams show a variety of topologies as a function of the disorder parameter $\textit{p}$. A comparison with recent results on the Blume-Capel model in random crystal field is discussed.
\end{abstract}

\begin{keyword}
Blume-Capel model \sep random crystal field \sep multicritical behavior \sep phase diagrams
\MSC[2010] 82B26 \sep 82B44 \sep 82B80
\end{keyword}


\end{frontmatter}

\linenumbers
%
%
\section{Introduction}
In recent years there has been increasing interest on the multicritical behavior of disordered systems. Special attention has been given to models with the inclusion of random fields, in the case of disordered magnetic systems, both for theoretical interest and also for its correspondence with the experimental results \cite{Belanger}.  Among those models, the Blume-Capel model \cite{Blume, Capel} and some of it extensions has received a lot of attention. The Blume-Capel is itself  an extension of the classical Ising model for spin-1 which takes into account the effect of a local crystal field anisotropy. Its phase diagram displays a line of continuous transition line which meets a first-order transition line which meet at a tricritical point \cite{BlumeEmeryGriffiths}. From the theoretical point of view a particularly interesting question is how such phase diagrams are changed under the effect of quenched randomness \cite{Harris, ImryMa, Berker, Hui}. Because of that, Kaufman and Kanner \cite{Kaufman} studied the Blume-Capel model under a random magnetic field and obtained a rich variety of phase diagrams.The effect  of random crystal field has been considered by a several authors \cite{Benyoussef,  Boccara89, Carneiro89, Carneiro90, Kaneyoshi, Maritan, Buzano, Borelli, Branco97, Lara98, Branco99, Salmon10, Albayrak, Gulpinar11, Yuksel, Gulpinar12, Lara12}. Besides the approach adopted, in some of these works the choice of the random crystal field distribution is also different. However, in all cases the phase diagrams display a rich behavior with the presence of critical and coexistence lines, as well as many multicritical points and re-entrant phenomena. In their recent work, Salmon and Tapia studied an infinite-range Blume-Capel under a quenched disorder crystal field following a superposition of two Gaussian distribution and classified the phase diagrams according to their topology \cite{Salmon10}. Recently, the effect of a special discrete random crystal field distribution was investigated by the pair approximation approach \cite{Lara12}. The same type of random crystal field distribution had already been investigated by the real-space renormalization-group approach, as well as by the mean-field approximation \cite{Branco97, Branco99}. However, as far as the results can be compared they lead to qualitatively different phase diagrams for low temperature. For instance, while the pair approximation predicts first-order transitions between the paramagnetic and ferromagnetic at zero temperature as in Figure 2 of \cite{Lara12}, the conclusion of the single-site mean-field approximation is that the ground state is always ordered according to Eq. (3) of \cite{Branco99}. Thus, we decided to investigate this point further by considering an exactly version of the Blume-Capel model under the random crystal field distribution considered by \cite{Branco97, Branco99, Lara12}. Besides, we are also interested in investigate the possible topologies for phase diagrams predicted by this sort of mean-field treatment along the lines of the continuous distribution started by\cite{Salmon10}. Finally, since the re-entrant phenomena in random spin-1 models has attracted some recent interest (see, for instante, \cite{daCosta} and references therein), our results may give some hint as to what expect in more disordered system as in the case of the Blume-Capel spin-glass under a random crystal field \cite{Ferrari}. 

This work is organized as follows. In Section 2 we introduce the Blume-Capel model under the presence of a random crystal field and obtain the basic equations. In Section 3 we present the obtained phase diagrams. Finally, we present our conclusions in Section 4.
%
%
\section{The Model}
Let us consider the infinite-range Blume-Capel model described by the following Hamiltonian:
\begin{eqnarray} 
{\cal H}=-\frac{J}{2N}\sum_{(i,j)}^{N} S_{i} S_{j} + \sum_{i=1}^N {\Delta}_{i} S_{i}^{2}\  ,
\label{eq:hamiltonian_BC}
\end{eqnarray}

\noindent
where $N$ is the number of spins and $S_i=-1, 0, +1$, for all sites $i = 1, \cdots, N$. The first sum runs over all pairs of spins $(i, j)$. The ferromagnetic coupling takes the form $J/N$ to account for the free energy extensivity. The random crystal fields ${\Delta}_i$ are quenched variables, independent and identically distributed according to the following probability distribution:
\begin{eqnarray} 
P({\Delta}_i;  D, p)= p\delta({\Delta}_i-D)+(1-p)\delta({\Delta}_i+D)\ .
\label{eq:probability_function}
\end{eqnarray}     

As far as we know, the above probability distribution was introduced by Branco and Boechat \cite{Branco97} and Branco \cite{Branco99} and has been recently considered by Lara \cite{Lara12}. The 
transformation ${\Delta}^{\prime}_i = ({\Delta}_i + D)/2$ leads to the probability distribution mostly used in the study of the Blume-Capel in discrete
random crystal field as, for instance, in \cite{Benyoussef, Carneiro90}, but also produces a slight change in the Hamiltonian. The general properties of the phase diagram should not depend on the particular form of the discrete random crystal field distribution. Thus, besides our interest in making comparison with known results (\cite{Branco97, Branco99, Lara12}), another reason for our choice of the probability distribution is the symmetry inherent in Eq.~(\ref{eq:probability_function}) which can be expressed by:

\begin{eqnarray} 
P({\Delta}_i; D, p) =  P({\Delta}_i; -D, 1-p).
\label{eq:sympd}
\end{eqnarray}  
Therefore, in order to determine phase diagrams for fixed values of $p$ it is sufficient to consider
the domain defined by $D \geq 0$ and $1/2 \le p \le 1$.

Using the replica method (see the Appendix for details), we obtain the free-energy density, in units of $J$:
\begin{eqnarray} 
 f(t,d,p;m)= &\displaystyle{\frac{1}{2}}m^2 - pt\ln\left[1+ 2\exp(-d/t)\cosh\left(m/t\right)\right] \nonumber \\ &-\left(1-p\right)t\ln \left[1+ 2\exp(d/t)\cosh\left(m/t\right)\right] ,
\label{eq:fed}
\end{eqnarray} 
where $t= k_BT/J$, $d=D/J$, and $m$ is the magnetization. The equation of state can be obtained
by taking the minimum of the above free-energy functional with respect to $m$, which leads to 
\begin{eqnarray}
m = \frac{2p \sinh\left(m/t\right)}{\exp(d/t)+2\cosh\left(m/t\right)} +\frac{2\left(1-p\right) \sinh\left(m/t\right)}{\exp(-d/t)+2\cosh\left(m/t\right)}.
\label{eq:mag}
\end{eqnarray}

The thermodynamic properties of the model is completely determined by  Eqs.~(\ref{eq:fed}) and (\ref{eq:mag}) which, in turn, reveals clearly the  symmetry expressed by Eq.~(\ref{eq:sympd}). For given values of $p$, $t$ and $d$ the physical solution corresponds to the global minima of the free-energy density. Thus, for a given value of $p$ we can determine the $d-t$ phase diagram. Eq.~(\ref{eq:mag})  always have a trivial solution corresponding to the paramagnetic phase \textbf{P}, with $m=0$. The corresponding paramagnetic free-energy density is given by                                                      
\begin{eqnarray} 
	f_P(t,d,p)= - pt\ln\left[1+ 2\exp(-d/t)\right] -\left(1-p\right)t\ln \left[1+ 2\exp(d/t)\right] .
	\label{eq:fpara}
\end{eqnarray}
Besides the paramagnetic solution, Eq.~(\ref{eq:mag}) may present distinct non-trivial solutions,  corresponding to different ferromagnetic phases.

Let us consider the ground state. For $d > 0$, the free-energy density $f_P$ for the paramagnetic solution becomes
\begin{eqnarray}
f_{0} \equiv f_P(t=0, d, p)= -\left(1-p\right)d.
\label{eq:f0}
\end{eqnarray}

Apart from the paramagnetic phase, we find two ferromagnetic solutions. The first type (\textbf{F1})  is characterized by $m_1=1$, with the free energy density given by:
 \begin{eqnarray}
 	f_{1} \equiv f(t=0,d,p;m=1) = -\frac{1}{2}+ \left(2p-1\right)d, \quad \mbox{for} ~ d<1 . 
 	\label{eq:f1}
 \end{eqnarray}
 The second type of ferromagnetic solution (\textbf{F2}) is given by $m_2= 1-p$, with the free energy density given by:
 \begin{eqnarray}
 	f_{2} \equiv f(t=0,d,p;m=1-p) = -\frac{1}{2}(1-p)^2 - (1-p)d , \quad \mbox{for} ~ d \ge 1-p .
 	\label{eq:f2}
 \end{eqnarray}
 
 From Eqs.~(\ref{eq:f0}) and (\ref{eq:f2}), we note that $f_2 \le f_0$ wherever the \textbf{F2} phase exists. Moreover, from the analysis of Eqs.~(\ref{eq:f0})$-$(\ref{eq:f2}) we find that the ground state 
 consists of the  \textbf{F1} phase for $d < d_0$, while for $d > d_0$ it corresponds to the \textbf{F2} phase. At zero temperature, $t= 0$, we determine a first-order transition between the  \textbf{F1} and \textbf{F2} phase at $d_0$ given by
 \begin{eqnarray}
 	d_0=1-\frac{1}{2}p .
 	\label{eq:d0}
 \end{eqnarray}
 Therefore, except for $p$ strictly equals to 1 the paramagnetic phase is never realized at zero temperature.

In general the $d-t$ phase diagrams for a given value of $p$ can be determined numerically from  Eqs.~(\ref{eq:fed}) and (\ref{eq:mag}). However, the stability of the paramagnetic phase can be determined analytically. From this analysis we can find critical frontiers as well as possible tricritical points. For this purpose, let us introduce the following parametrization:
\begin{eqnarray}
a=\mbox{exp}(d/t) .
\label{eq:parametrization}
\end{eqnarray}
Nearby a continuous transition from ferromagnetic to paramagnetic phase, we consider a small magnetization $m \simeq 0$ and write a Landau-like expansion for the free energy density:
\begin{eqnarray}
f(t, d, p; m) = A_0 + A_{2}m^2 + A_{4}m^4 + A_{6}m^6 + \cdots .
\label{eq:Landau}
\end{eqnarray}
The coefficient $A_0$ corresponds to $f_P(t,d,p)$ given by Eq.~(\ref{eq:fpara}), while the remaining coefficients are given by:

\begin{eqnarray}
A_2 & =& \frac{1}{2} -\frac{p}{2t} q - \frac{(1-p)}{2t} r  ,
\label{eq:f_2} \\
A_4 &=& -\frac{p}{24t^3} (1-3q) q -\frac{(1-p)}{24t^3} (1-3r) r,
\label{eq:f_4} \\
A_6 &=& -\frac{p}{720 t^5} (1 - 15q + 30q^2) q  -\frac{(1-p)}{720 t^5} (1 - 15r + 30r^2) r, 
\label{eq:f_6}
\end{eqnarray}

\noindent
where $q$ and $r$ given by
\begin{eqnarray}
q = \frac{2}{2+a} , \quad r = \frac{2a}{2a+1}.
\label{eq:q&r}
\end{eqnarray}
These new parameters $q$ and $r$ are not independent and can be interpreted as the density of spins $S_i = \pm 1$ in the paramagnetic phase for the pure cases $p=1$ and $p=0$, respectively.

A continuous transition line from the ferromagnetic to paramagnetic phase is given by 
$$ A_2 = 0, \quad \mbox{while} ~~ A_4 > 0 . $$
From ~(\ref{eq:parametrization}), one has the following expression for the critical line:
\begin{eqnarray}
t=\frac{2p}{2+a}+\frac{2\left(1-p\right)a}{2a+1} ,
\label{eq:tcrit}
\end{eqnarray}
which is valid as far as $A_4 > 0$.

As in the pure case $p =1$, we can find tricritical point when 

$$ A_2 = 0, ~~ A_4 = 0, \quad \mbox{while} ~~ A_6 > 0 . $$

Since the coefficients $A_2, A_4$ and $A_6$ are functions of $t$, $d$ (trough $a$) and $p$, we find a threshold for the tricriticallity as $ A_2 = A_4 = A_6 = 0 $, which corresponds to a multicritical point when $A_8 > 0$ (sometimes called the last tricritical point). Thus, we determine the value 
\begin{eqnarray}
 p^{\star} = 0.978400\dots ,
 \label{eq:pstar}
\end{eqnarray}
such that for $p^{\star} < p \le 1$ there are tricritical points in the $d-t$ phase diagrams. 

For $p < p^{\star}$ the tricritical behavior disappears leaving room to other critical or multicritical behavior as the appearance of critical endpoints. Finally, we can also determine a value of $p = p_r = 0.982811\dots $ such that the tricritical point coincides with the maximum of the critical line $d = d(t)$ implicitly determined from Eq.~(\ref{eq:tcrit}). Thus, for $p^{\star} \le p < p_r$ there are  re-entrant phenomena from the ferromagnetic to the paramagnetic phase associated with the tricritical behavior. In the following section we present our results in terms of the $d-t
$ phase diagrams.

\section{Phase Diagrams}

The phase diagrams were determined by numerically finding the global mininum of the free energy density given by Eq.~(\ref{eq:fed}). In general the $d-t$ phase diagrams consists of continuous transition as well as first-order transitions lines and some special, or multicritical, points. The pure case, which corresponds to $p=1$, is well-known to display both first-order as well as continuous transition lines 
separating the ferromagnetic and paramagnetic phase. These lines meet at the so-called tricritical point \cite{BlumeEmeryGriffiths}. Since we are interested in the disordered case, we will no longer discuss the phase diagram for the pure case. As in similar models, the tricritical behavior is affected by the randomness. Besides, the randomness may cause the appearance of ordered critical point, which is the end of the coexistence line between two ordered phases, as well as critical endpoint, which is the end of critical line on a coexistence curve. Following \cite{Salmon10}, we use the following convention in our phase diagrams:
\begin{itemize}
	\item continuous transition or critical line: continuous line;
	\item first-order transition line: dotted line;
	\item tricritical point: located by a black circle;
	\item ordered critical point: located by an asterisk;
	\item critical endpoint: located by a black triangle;
\end{itemize}

Depending on the value of $p$ we found essentially three topologically distinct phase diagram in the $d-t$ plane. 

The phase diagrams belonging to Topology I occurs for $1/2 \le p < p_{l} = 0.926277\dots$. They present
a continuous transition line between the paramagnetic and ferromagnetic phases, as displayed in Figure 1.
For low temperatures there is a first-order order transition line separating distinct ferromagnetic phases \textbf{F1} and \textbf{F2}. This line ends at an ordered critical point. 

The Topology II phase diagrams appear for $ p_{l} < p < p^{\star}$. Their general behavior is ilustrated in Figure 2. The paramagnetic phase is separated from the ordered (ferromagnetic) phases by first-order and continuous transition lines. For low temperature and $d \gtrsim 1-p/2$ there is a critical line between the paramagnetic and the ferromagnetic
\textbf{F2} phase. Also, for low temperatures and $d \cong 1-p/2$ there is a first-order transition between the \textbf{F1}$-$\textbf{F2} phases. These two lines meet at a critical endpoint with a second first-order transition line between the paramagnetic \textbf{P} and the \textbf{F1} phase. This new line ends at another critical endpoint common to the critical line between the paramagnetic and ferromagnetic phases. Finally, this second critical endpoint belongs to a second first-order line between the ferromagnetic phases \textbf{F1} and \textbf{F3}. This kind of topology is also characterized by re-entrant effects for intermediate temperatures and $d$ slight greater than $1-p/2$.

The third type of phase diagrams defines the Topology III. They are obtained for $p^{\star} < p < 1$. The intermediate to low temperature regime is quite similar to Type II phase diagram. However, the paramagnetic and ferromagnetic \textbf{F1} phase are separated by a coexistence line at intermediate temperatures and a continuous transition line for higher temperatures. These two lines meet together at a tricritical point. In this case, for a small $p$-window we have reentrant phenomena ($p^{\star} < p <p_r$). A typical case of Topology III phase diagram is shown in Figure 3.

An important consequence of the random anisotropy distribution used in the present work is that the paramagnetic phase cannot be realized at zero temperature for $p$ strictly less than 1. Thus, in our case there is no phase diagram corresponding to topology I described by \cite{Salmon10}. In particular, our findings do not reproduce the structure of the phase diagrams obtained by \cite{Lara12} who find stable paramagnetic phases at zero temperature as can be seen in their Figures 1 and 2. Also, our low temperature results are at variance with those obtained by real-space renormalization group analysis \cite{Branco97} for low dimensional systems. Perhaps these disagreements are due the mean-field character of our infinite-range model. We plan to investigate this further in future works.

\begin{figure}[!htbp]
	\centering
		\includegraphics[scale=0.38]{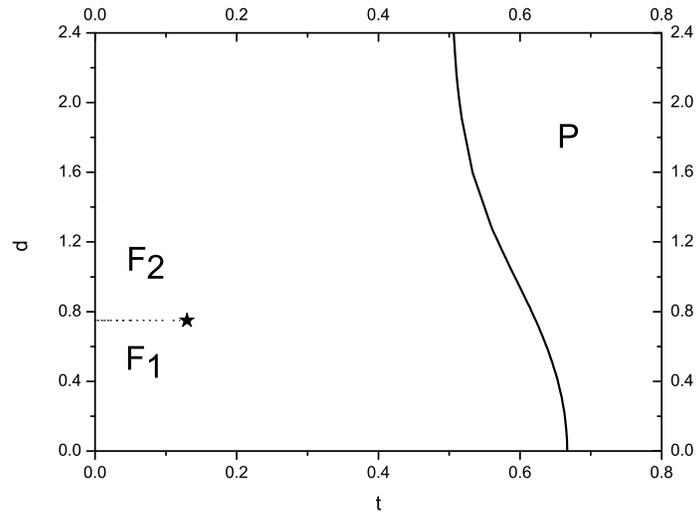}
	\caption{$d-t$ phase diagram for $p=0.5$, corresponding to Topology I, displaying a continuous
		transition and a first-order transition line, as well an ordered critical point.}
\label{fig:grafico_p05}
\end{figure}

\begin{figure}[!htbp]
	\centering
		\includegraphics[scale=0.38]{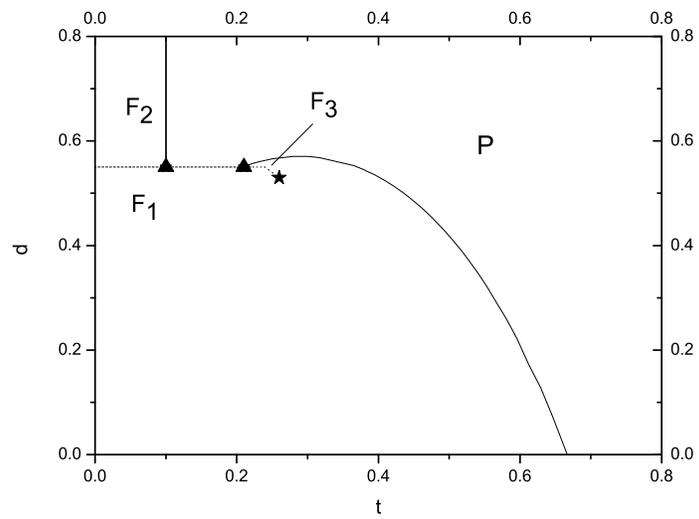}
	\caption{$d-t$ phase diagram for $p=0.9$, corresponding to Topology II, with three
		first-order transition lines (coexistence of \textbf{F1-F2}, \textbf{F1-P} and \textbf{F1-F3} phases), and two critical lines separating the paramagnetic from the
		ordered ferromagnetic. A re-entrant effect is clearly noted in this phase diagram. It is also shown an ordered critical point and two critical endpoints.}
\label{fig:grafico_p09}
\end{figure}

\begin{figure}[!htbp]
	\centering
		\includegraphics[scale=0.38]{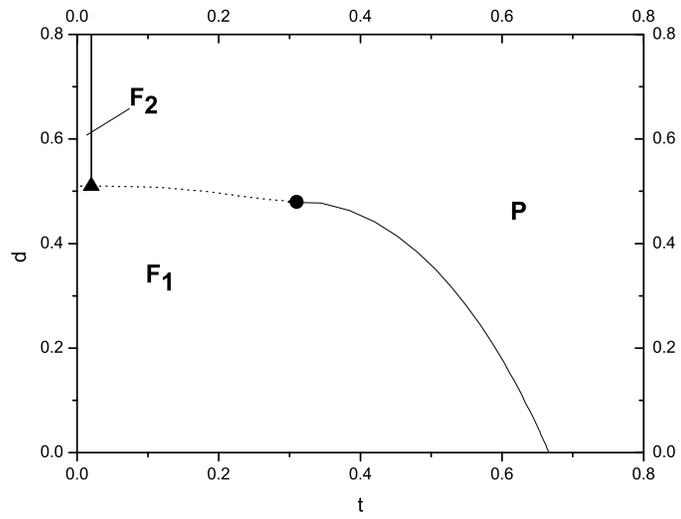}
	\caption{$d-t$ phase diagram  for $p=0.98$, corresponding to Topology III and displaying
		two first-order transition lines (coexistence of \textbf{F1-F2} and \textbf{F1-P} phases), and two critical lines separating the paramagnetic from the ferromagnetic phases \textbf{F1} and \textbf{F2}. In this case we have a tricritical point and a critical endpoint.}
\label{fig:grafico_p098}
\end{figure}

 \newpage
\section{Conclusions}
In this work we studied the infinite-range Blume-Capel model under a quen\-ched random crystal field.  The exact free energy density was determined by means of the replica method. A variety of rich phase diagrams, displaying critical and coexistence lines as well as ordered critical, tricritical and critical end points was obtained. These phase diagrams can be classified according to three classes, designated as topologies I, II and III, with increasing values of the parameter $\textit{p}$ which is a measure of the crystal field randomness. Topology I is characterized by a continuous critical line separating the high temperature paramagnetic phase from the ordered ferromagnetic phase. For low temperatures we have two ferromagnetic \textbf{F1} and \textbf{F2} phases, which are separated by a first-order transition line ending at an ordered critical point. For Topology II the paramagnetic phase is separated from the ordered ferromagnetic phases either by critical or coexistence lines. For low temperatures the critical \textbf{P}$-$\textbf{F2} line meet two coexistence lines (\textbf{F1}$-$\textbf{F2} and \textbf{P}$-$\textbf{F1})  at a critical endpoint. On the other hand, for intermediate temperatures the critical line between the paramagnetic and ferromagnetic phases ends at a second critical endpoint. This critical endpoint is also the terminus of the \textbf{P}$-$\textbf{F1} coexistence line as well as of the  \textbf{F1}$-$\textbf{F3} coexistence lines which, in turn, ends at a common critical point.  The third class of phase diagram belongs to Topology III. It is marked by the disappearance of the ferromagnetic \textbf{F3} found in the previous case so that the critical \textbf{F1}$-$\textbf{F3} point merge with the  \textbf{P}$-$\textbf{F1}$-$\textbf{F3} critical endpoint, giving rise to a tricritical point which is the meeting point of the critical \textbf{P}$-$\textbf{F1} boundary with the \textbf{P}$-$\textbf{F1} coexistence line. In case of Topology II we always have the possibility of reentrant phase transitions, whereas for Topology III reentrant phenomena can only occur for a narrow window in terms of $p$. It remains to be seen whether these effects are present in more realistic short-range models.

\section*{Acknowledgement}
The partial financial support from Capes (Brazilian agency) is acknowledged.
\appendix
\section*{Appendix}
\numberwithin{equation}{section}
Apart from terms which do not contribute in the termodynamic limit we can rewrite the Hamiltonian (\ref{eq:hamiltonian_BC}) as
\begin{eqnarray}
{\cal H}=-\frac{J}{2N}\left(\sum_{i=1}^{N} S_{i}\right)^2 + \sum_{i=1}^N \Delta_{i} S_{i}^{2}\  .
\label{eq:hamiltonian2}
\end{eqnarray}
Since we are dealing with a quenched system, the free energy is given by 

\begin{equation}
F = -k_BT \langle \ln Z \rangle ,
\label{eq:fe}
\end{equation}
where $\langle ...\rangle$ denotes the average over the disorder given by Eq.~(\ref{eq:probability_function}).
In the thermodynamic limit the free energy density is given by
\begin{equation}
f = \lim_{N \to \infty} \frac{F}{N} = -k_BT\lim_{n \to \infty} \frac{1}{N} \langle \ln Z \rangle .
\label{eq:fedf}
\end{equation}
In order to compute the averaged free energy (\ref{eq:fe}) we make use of the replica method \cite{Dotsenko, Nishimori} based on the identity
\begin{equation}
\ln Z = \lim_{n \to 0} \frac{1}{n} (Z^n - 1) .
\label{eq:replicatrick}
\end{equation}
Writing
\begin{equation}
Z^n = \prod_{\alpha=1}^n \tr \exp\left[\frac{\beta J}{2N}\left(\sum_{i=1}^{N} S_{i}^{\alpha}\right)^2 -\beta \sum_{i=1}^N \Delta_{i} (S_{i}^{\alpha})^{2} \right] ,
\end{equation}
and using the identity

\begin{equation}
e^{\frac{\beta J}{2N}\left(\sum_{i=1}^{N} S_{i}^{\alpha}\right)^2} = \int_{-\infty}^{\infty}e^{-\frac{1}{2} N\beta J m_{\alpha}^2 + \beta J m_{\alpha} \sum_{i} S_{i}^{\alpha}} \frac{d\hat{m}_{\alpha}}{\sqrt{2\pi}}, 
\end{equation}
where $\hat{m}_{\alpha} = \sqrt{N \beta J}m_{\alpha}$ and taking into account that disorder is local for each site $i$ we obtain

\begin{equation}
\langle Z^n \rangle = \prod_{\alpha=1}^n\int_{-\infty}^{\infty}  \frac{d\hat{m}_{\alpha}}{\sqrt{2\pi}} e^{-\frac{1}{2}N\beta J m_{\alpha}^2 + N \langle \ln \tr \exp (\beta Jm_{\alpha}S^{\alpha} - \beta \Delta (S^{\alpha})^2) \rangle} 
\label{eq:znavg}
\end{equation}
In the large $N$ limit the integral in Eq.~(\ref{eq:znavg}) is dominated by the maximum and can be evaluated by the steepest descent method. Since we have equivalent and nointeracting replica the saddle point is given by $m_{\alpha} = m$ for any replica $\alpha$. Thus we can finally obtain, for $ N >> 1 $, the asymptotic expression
\begin{equation}
\langle Z^n \rangle \cong \exp \left( -N n \min_{\{{m}\}} f(m) \right),
\label{eq:zn_saddlepoint}
\end{equation}
where
\begin{eqnarray}
f(m)= \displaystyle{\frac{\beta J}{2}}m^2 - \langle \ln \tr e^{\beta J m S - \beta \Delta S^2} \rangle ,
\end{eqnarray}
from which we obtain the free energy in units of $J$ (this is equivalent to taking $J=1$) given by Eq. (\ref{eq:fed}).

\newpage
%

%
%
\end{document}